\documentclass[aps,prd,twocolumn,showpacs,groupedaddress,nofootinbib]{revtex4}
\usepackage{graphicx}
\usepackage{color}
\begin{document}

\title{$f(2010)$ in  Lattice QCD}

\author{Mushtaq Loan$^{a}$\footnote{Corresponding author}, Zhi-Huan Luo$^{b}$ and Yu Yiu Lam$^{c}$}
\affiliation{$^{a}$ International School, Jinan University, Huangpu Road West, Guangzhou 510632, P.R. China\\
$^{b}$ Department of Applied Physics, South China Agricultural
University, Wushan Road, Guangzhou, 510642, P.R. China\\
$^{c}$ Department of Physics, Jinan University, Huangpu Road West,
Guangzhou 510632, P.R. China}
%\date{\today}
\date{May 27, 2009}
\begin{abstract}
We present a search for the possible $I(J^{P})=0(2^{+})$
tetraquark state with $ss{\bar s}{\bar s}$ quark content in
quenched improved anisotropic lattice QCD. Using various local and
non-local interpolating fields we determine the energies of
ground-state and second ground state using variational method. The
state is found to be consistent with two-particle scattering
state, which is checked to exhibit the expected volume dependence
of the spectral weights.  In the physical limit, we obtain for the
ground state, a mass of $2123(33)(58)$ MeV which is higher than
the mass of experimentally observed $f(2010)$. The lattice
resonance signal obtained in the physical region does not support
a localized $J^{P} =2^{+}$ tetraquark state in the pion mass
region of $300 - 800$ MeV. We conclude that the $4q$ system in
question appears as a two-particle scattering state in the quark
mass region explored here.
\end{abstract}

\maketitle

\section{INTRODUCTION}

The concept of multi-quark hadrons has received revived interest
due to the narrow resonances in the spectrum of states. Recently,
several new particles were experimentally discovered and confirmed
as the candidates of multi-quark states.  These discoveries are
expected to reveal new aspects of hadron physics. Among these
discoveries, the tetra-quark systems are also interesting in terms
of their rich phenomenology, in particular for mesons which still
remain a most fascinating subject of research. The $4q$ states are
interesting in terms of the recent experimental discoveries of
$X(3872)$ \cite{Belle03,CDF04,D004}, $Y(4260)$ \cite{BABAR04} and
$D_{s}(2317)$ \cite{BABAR03,Belle03b}, which are expected to be
tetra-quark candidates.

The Particle Data Group  lists $2$  tensor mesons with masses in
the range $1.9 - 2.2$ $\mbox{GeV}/c^{2}$ and considers them as
well-established. The  $2^{++}$ candidates, $f_{2}(1950)$
\cite{BES00,Abe04,WA00} and $f_{2}(2010)$ \cite{BNL88} are
isosinglet. The relevant channels of decay are $K{\bar K}$ and
$\eta\eta$ for the $f_{2}(1950)$ and $\phi\phi$ and $K{\bar K}$
for $f_{2}(2010)$. Due to their $K{\bar K}$ decay, one would
expect $f_{2}(1950)$ and $f_{2}(2010)$ are very likely one state;
the mass shift could be a measurement error or could be caused by
the $K{\bar K}$ threshold. However, the results for $f_{2}(2010)$
favour an intrinsically narrower state, strongly coupled to
$\phi\phi$ and weakly coupled to the other channel for allowed
$s$-wave decays. Following the recent re-analysis of the BNL data
\cite{Longacre04} we discuss the state $f_{2}(2010)$ as a
$s^{2}{\bar s}^{2}$ state.

The multi-quark states have been investigated in lattice QCD
studied with somewhat mixed results
\cite{Lasscock,Oikharu,luo07,Loan08a}. At the present status of
approximations, lattice QCD seems to provide a trustworthy guide
into unknown territory in tetra-quark hadron physics
\cite{Loan08b,Fukugita,Mark,Gupta,Sharpe,Hideo,Alexandrou,Fumiko,Mathur07,
Ting06,Sasa09}. Using the quenched approximation, and discarding
quark-antiquark annihilation diagrams, we construct $s^{2}{\bar
s}^{2}$ sources from multiple operators.  Note that we are working
in quenched approximation which in principal is unphysical.
However, previous lattice results  on masses and decay constants
turn out to be in good agreement with experimental values
\cite{Alford00}. This seems to suggest that it is plausible to use
quenched lattice QCD to investigate the mass spectra. We exclude
the processes that mix $q{\bar q}$ and $q^{2}{\bar q}^{2}$ and
allow the quark masses to vary from small to large values. In the
absence of quark annihilation, we do not expect any mixing of
$q^{2}\bar{q}^{2}$ with pure glue. Thus we can express the
$q^{2}\bar{q}^{2}$ correlation functions in terms of a basis
determined by quark exchange diagrams only (ignoring the single,
double and annihilation diagrams among Wick's contractions).
Another important question is whether the interpolating operator
one uses has a significant overlap with the state in question. To
construct an interpolating field which has significant overlap
with the $4q$ system, we adopt the so-called variational method to
compute $2\times 2$ correlation matrix from two different
interpolating fields  and from its eigenvalues we extract the
masses. Thus, assuming that the quenching uncertainties do not
effect our conclusions dramatically, we investigate the optimized
correlation function  and use it to examine lowest-lying
tetra-quark resonance as $f_{2}(s{\bar s}s{\bar s})$ states in the
spectrum of $2\times 2$ correlation matrix.

\section{Lattice study for the $f_{2}(ss{\bar s}{\bar s})$}
The simplest local interpolators can be written in terms of
colour-singlet configuration of a product of colour-neutral meson
interpolation fields.  We propose a non-$\phi\phi$ interpolating
field to extract the $f_{2}(ss{\bar s}{\bar s})$ tetraquark state.
This choice is designed to maximize the possibility to observe
attraction between tetraquark constituents at relatively large
quark masses. With a $\phi\phi$ operator it is possible that there
is a  small amount of the compact $4q$ component in the two-body
interpolating field since the interpolator may contain a large
contamination of $\phi\phi$ scattering states. We adopt the
simplest non-$\phi\phi$- type interpolator of the form
\begin{eqnarray}
O_{1}(x) & =&  \frac{1}{2}\left[\bigg({\bar q^{a}}_{\alpha}(x)
(\gamma_{i})_{\alpha\beta}q^{a}_{\beta}(x)\bigg) \bigg({\bar
Q^{b}}_{\lambda}(x)(\gamma_{j})_{\lambda\sigma}
Q^{b}_{\sigma}(x)\bigg) \right.
\nonumber\\
& & \left. - \bigg(q\leftrightarrow Q, {\bar q} \leftrightarrow
{\bar Q}\bigg)\right], \label{eqn01}
\end{eqnarray}
with spin $I(J^{P})=0(2^{+})$. In the nonrelativistic limit the
above non-two state particle can not be decomposed into
$\phi\phi$. Thus the $4q$ state can be singled out as much as
possible and the results are less biased by the contamination of
two-state scattering states.

The other type of interpolating field is one in which quarks and
anti-quarks are coupled into a set of diquark and antidiquark,
respectively and has the form
\begin{equation}
O_{2}(x)  = \epsilon_{abc}\left[q^{T}_{b}C\Gamma
Q_{c}\right]\epsilon_{ade}\left[{\bar q}_{d}C\Gamma {\bar
Q}^{T}_{e}\right]. \label{eqn02}
\end{equation}
Accounting for both colour and flavour antisymmetry, possible
$\Gamma$s are restricted within $\gamma_{5}$ and $\gamma_{i}$. For
$\Gamma = \gamma_{5}\gamma_{i}$ ($i=1,2,3$), the above diquark
operator transforms like $J^{P}=1^{-}$. For concreteness, we
simulate the flavour combination $[ss]$ and $[{\bar s}{\bar s}]$.

To extract energies $E_{n}$ of $s^{2}{\bar s}^{2}$ We compute the
$2\times 2$ correlation matrix
\begin{equation}
C_{ij}(t) =\langle \sum_{ \vec x}\mbox{tr}\left[\langle
(O_{i})({\vec x},t){\bar O}_{j})({\vec
0},0)\rangle_{f}\right]\rangle_{U},
\label{eqn04}
\end{equation}
where the trace sums over the Dirac space, and the subscripts $f$
and $U$ denote fermionic average and gauge field ensemble average,
respectively. Following \cite{Lasscock,Loan08b,Alexandrou05} we
solve the eigenvalue equation
\begin{equation}
C(t_{0})v_{k}(t_{0})=\lambda_{k}(t_{0})v_{k}(t_{0})
\label{eqn05}
\end{equation}
to determine the eigenvectors $v_{k}(t_{0})$. We use these
eigenvectors to project the correlation matrices to the space
corresponding to the $n$ largest eigenvalues $\lambda_{n}(t_{0})$
\begin{equation}
C_{ij}^{n}(t) = (v_{i},C(t)v_{j}), \hspace{1.0cm} i,j=1, \cdots ,n
\label{eqn06}
\end{equation}
and solve the generalised eigenvalue problem equation for the
projected correlation matrix $C_{ij}^{n}$. The resulting
large-time dependence of the eigenvalues $\lambda_{n}(t)$ allows a
determination of ground and excited-state energies. The mass can
be extracted by a hyperbolic-cosine fit to $\lambda_{n}(t)$ for
the range of $t$ in which effective mass
\begin{equation}
M_{eff}(t) = \ln\left[\frac{\lambda (t)}{\lambda (t+1)}\right]
\label{eqn07}
\end{equation}
attains a plateau. In order to show the existence or absence of
the signature of tetraquark resonance on lattice, we establish
lowest and  as well as the second-lowest energy levels for our
$4q$ system.

Using a tadpole-improved anisotropic gluon action
\cite{Morningstar99}, we generate quenched configurations on two
lattice volumes $16^{3}\times 64$ and $16^{3}\times 80$ (with
periodic boundary conditions in all directions). After discarding
the initial sweeps, a total of 200 configurations are accumulated
for measurements at $\beta=4.0$ Quark propagators are computed by
using a tadpole-improved clover quark action on the anisotropic
lattice \cite{Okamoto02}. All the coefficients in the action are
evaluated from tree-level tadpole improvement.

The bare mass of the strange quark is determined by extracting the
mass of the vector meson $M_{\phi}$. At $m_{q}a_{t} =0.066$, we
obtain $\kappa_{t}= 0.2404$, which produces a mass for the $\phi$
of $1.237(2)$ in lattice units. Using the mass of the nucleon in
the chiral limit, we find that the ratio $M_{\phi}/M_{N}$ at the
chiral limit is $1.059\pm 0.014$, which is in good agreement with
the physical ratio of $1.087$. This verifies that the strange bare
quark mass of $0.07$ used is very close (within $3\%$) to the
physical strange quark mass. The quark propagators are then
computed at seven values of the hopping parameter $k_{t}$ which
cover the strange quark mass region of $m_{s} <m_{q} <2m_{s}$,
i.e., $a_{t}m_{q} = 0.07, 0.075, 0.08, 0.09, 0.105, 0.115, 0.12$.
Inspired by the good agreement of the ratio with the experimental
value, the scale was set alternatively by  $M_{\phi}/M_{N}$. Using
the experimental value $938$ MeV for the nucleon mass, the spacing
of our lattice is $a_{s} = 0.473(2)$ fm.

\section{Results and discussion}

\begin{figure}[!h]
\scalebox{0.45}{\includegraphics{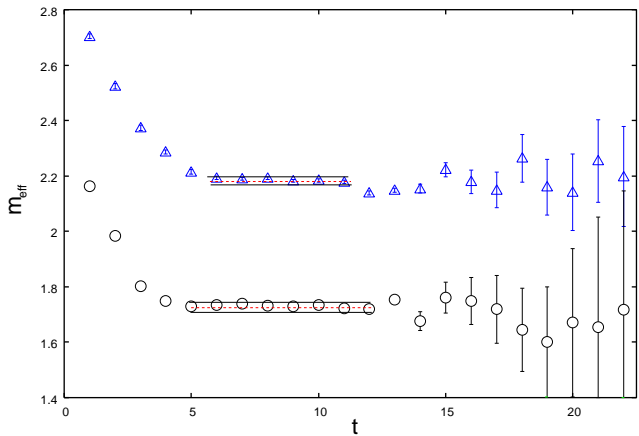}} \caption{ \label{fig1}
Effective mass of the $I(J^{})=0(2^{+})$ colour-singlet
lowest-lying ground state. The data correspond to $m_{\pi}\simeq
361$ MeV (triangles) and $824$ MeV (circles).}
\end{figure}

Figs. \ref{fig1}  illustrate the two lowest energy levels
extracted by fitting the effective masses over appropriate $t$
intervals.  The ground state eigenvalues show a conventional
time-dependence near $t\simeq T/2$ and hence the mass can be
accurately extracted using Eq. (\ref{eqn07}). We choose one ``best
fit" which is insensitive to the fit range, has high confidence
level and reasonable statistical errors. We then confirm this by
looking at the plateau region of the correlator. Statistical
errors of masses are estimated by the jackknife method and the
goodness of the fit is gauged by the $\chi^{2}/N_{DF}$, chosen
according to criteria that $\chi^{2}/N_{DF}$ is preferably close
to $1.0$.

The effective mass  is found to be stable using different values
of $t$ in Eq. (\ref{eqn07}), which suggests that the ground state
in question is correctly projected. Suppressing any data point
which has error larger than its mean value, the possible plateau
is seen in the region $5\leq t\leq 12$ with reasonable errors,
where the single-state dominance is expected to be achieved.
Fitting the effective mass in the window $t=6 -11$ is found to
optimize the $\chi^{2}/N_{DF}$. To avoid the clutter in  Fig.
\ref{fig1} we do not show the points at larger $t$ values which
have larger error bars, and have little or no influence on the
fits. The best fit curve to the $4q$ data has $\chi^{2}/N_{DF}=
0.87$. The results for the masses corresponding to the various
values of the hopping parameter $\kappa_{t}$ are tabulated in
Table \ref{tab1}.

\begin{table}[!h]
\caption{ \label{tab1} The masses of the  $4q$, Kaon and $\phi$
states, in the lattice units, for various values of $\kappa_{t}$.}
\begin{ruledtabular}
\begin{tabular}{ccccccc}
$\kappa_{t}$ &  $M_{4q}$ & $M_{4q}^{*}$ &  $M_{K}$ & $M_{\phi}$ \\
\hline
0.2410   & 2.187(4)  & 2.223(6)  & 0.579(2)  & 1.029(3)  \\
0.2420   & 2.055(7)  & 2.083(9)  & 0.507(5)  & 0.973(6)  \\
0.2435   & 1.951(13) & 1.973(11) & 0.475(8)  & 0.926(11)  \\
0.2440   & 1.812(19) & 1.825(21) & 0.440(5)  & 0.860(15) \\
0.2450   & 1.724(24) & 1.751(37) & 0.417(13) & 0.822(28)  \\
0.2455   & 1.655(27) & 1.673(44) & 0.401(14) & 0.787(17) \\
0.2462   & 1.591(23) & 1.606(49) & 0.383(19) & 0.759(24)
\end{tabular}
\end{ruledtabular}
\end{table}

\begin{figure}[!h]
\scalebox{0.45}{\includegraphics{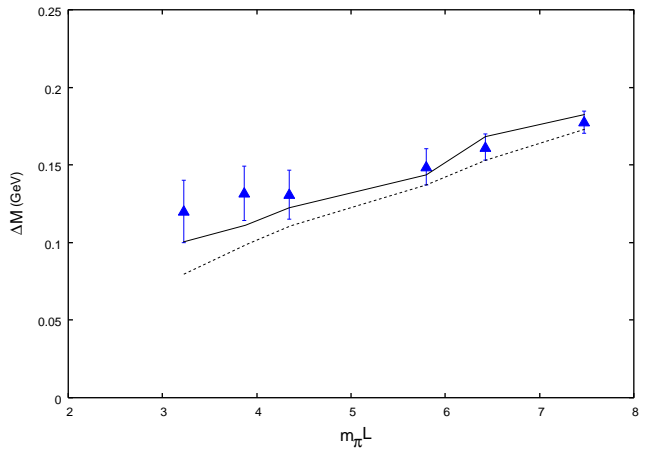}} \caption{
\label{fig2} The energy shift of the lowest $I(J^{P})=0(2^{+})$
state as a function of $m_{\pi}L$. The solid and dashed lines
correspond quenched one-loop chiral perturbation results for
$\delta = 0.12$ and $0.15$, respectively.}
\end{figure}

To interpret the ground state in terms of signatures of a lattice
resonance, we look at three possible scenarios. First, we extract
the mass splitting between the tetraquark $0(2^{+})$ and the
noninteracting $\phi +\phi$ two-particle state and compare our
results to that derived in quenched chiral perturbation
theory\footnote{Since we are using the quenched approximation, the
extraction of energy shift in a finite box using full QCD one-loop
chiral perturbation theory is not applicable}. Fig. \ref{fig2}
shows the  mass difference $\Delta M =M_{4q}-2M_{\phi}$, together
with the quenched one-loop energy shift in the finite box
\cite{Bernard96}, as a function of $m_{\pi}L$ for the lowest $4q$
state from $16^{3}\times 64$ lattice in our calculation. We obtain
the results for one-loop energy shift by interpolating the
coefficients $A_{0}(m_{\pi}L)$ and $B_{0}(m_{\pi}L)$ listed in
Ref. \cite{Bernard96} for the range of $m_{\pi}L$ appropriate for
our calculation on $16^{3}\times 64$ lattice for $\delta =0.12$
and $0.15$.

We see clearly that the masses derived for the tetraquark state
are consistently higher than the lowest two-particle state. The
mass difference is over $ 100$ MeV at small quark masses and
weakly dependent on $m_{\pi}$L.  The positive mass difference
observed in this range of pion mass suggests that the observed
signal is unlikely to be a tetraquark. We also notice that our
data are reasonably consistent with one-loop quenched perturbation
results \cite{Bernard96} for $m_{\pi}L \geq 4.3$ for $\delta =
0.12$  and $0.15$. It is interesting to note that our results are
consistent with quenched one-loop results despite the fact that
the disconnected contributions were not inserted in our
calculation. This implies that disconnected correlator has very
small or  negligible contribution than the connected one at
several time separations.

To confirm or discard  the signature observed in Fig. \ref{fig2},
we examine the second scenario, i.e., the volume dependence of the
spectral weight of these states. Theoretically, if the state is a
genuine resonance, then its spectral weight should be almost
constant for any lattices with the same lattice spacing. On the
other hand, if it is a two-particle scattering state, then its
spectral weight has an explicit $1/V^{3}$ dependence
\cite{Mathur04}. In the following, we shall use the ratio of the
spectral weights on two spatial volumes $16^{3}$ and $20^{3}$ to
discriminate whether the hadronic state in question is a resonance
or a scattering state.

\begin{figure}[!h]
\scalebox{0.45}{\includegraphics{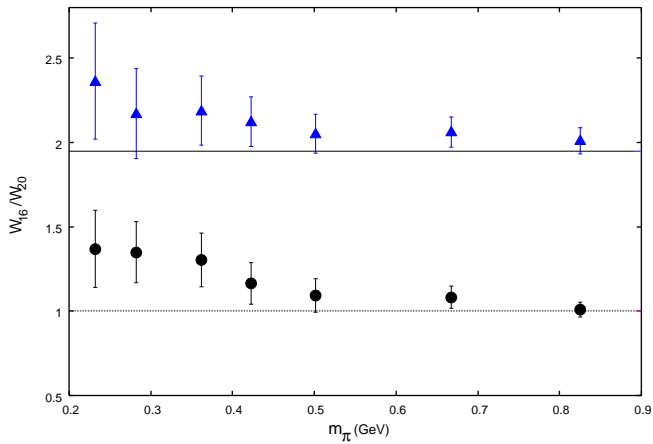}} \caption{ \label{fig3}
Spectral weight ratio $W_{16}/W_{2}$ as a function of
$m_{\pi}^{2}$ for the lowest state (solid circles) and next lowest
state (solid triangles).}
\end{figure}

Fig. \ref{fig3} shows the ratio ($R=W_{16}/W_{20}$) of spectral
weights of the lowest state  and second-lowest state, extracted
from the time-correlation function of variational matrix as a
function of $m_{\pi}^{2}$. Since our two lattice sizes are
$16^{3}\times 64$ and $20^{3}\times 80$, the spectral weight ratio
for a two-particle state should be $W_{16}/W_{20}=V_{20}/V_{16} =
1.95$. We see that the ratio $R$ for the lowest state clusters
around $1.0$ for $m_{\pi} \in [0.5,0.8]$, which implies that there
exists a $2^{+}$ resonance with quark contents $(ss{\bar s}{\bar
s})$.

On the other hand, for smaller quark masses, $R$ begins to deviate
from $1.0$ with larger errors, suggesting that this state is a
scattering state. Since none of our operators has scalar meson
component, the possibility that this might be due to quenching
effects at smaller quark masses is highly unlikely. Thus one can
safely ignore the possibility of $R$ being consistent with $1.0$
if one incorporates internal quark loops with larger volumes. This
type of flip-flop between the $4q$ state and the two-$\phi$ state
might be a flux -tube recombination between two $\phi$ at some
diquark and internal quark separations. This can be verified by
analyzing the $4q$ potential of the tetraquark system. We do not
intend to pursue such an analysis here since this is not the focus
of our present study. The spectral weight ratio of the first
excited state turns out to be consistent with $1.95$, confirming
our speculation that it is two-particle scattering state. The two
states  are reasonably well separated compared to the decay width
of $f(2010)$.

\begin{figure}[!h]
\scalebox{0.45}{\includegraphics{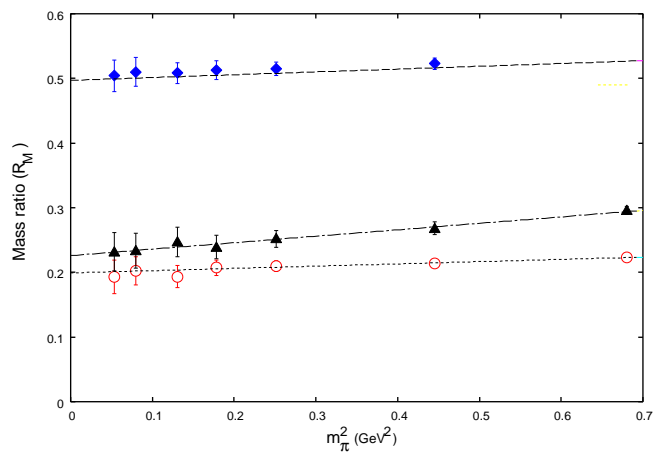}}
\caption{\label{fig4} Extrapolation of the mass ratio  $\Delta
M/M_{K}$ for lowest state (open circle) and second lowest state
(solid triangles) to the physical limit at $a_{s}=0.473$ fm. Also
are shown the mass ratio $M_{K}/M_{\phi}$ (solid diamonds). The
dashed lines are the linear fits in $m^{2}_{\pi}$ to the data.}
\end{figure}

Finally, the  mass differences extracted can be extrapolated to
the physical limit, which is the next important issue
\cite{Derek04}. Since quenched spectroscopy is quite reliable for
mass ratio of stable particles, it is physically even more
motivated to extrapolate mass ratios rather than masses or mass
differences. This allows for the cancellation of systematic errors
since the hadron states are generated from the same gauge field
configurations and hence systematic errors are strongly
correlated. We use a set of data points with smallest
$m_{\pi}^{2}$ to capture the chiral log behaviour. Fig. \ref{fig4}
collects and displays the resulting mass ratios, illustrated in
Table \ref{tab2},  extrapolated to the physical limit using linear
and quadratic fits in $m_{\pi}^{2}$. The difference between these
two extrapolations gives some information about systematic
uncertainties in the extrapolated quantities. Performing such
extrapolations to mass ratios, we adopt the choice which shows the
smoothest scaling bahaviour for the final value, and use others to
estimate the systematic errors.

The data at smallest five quark masses behave almost linearly in
$m_{\pi}^{2}$ and both the linear and quadratic fits essentially
gave the identical results. The contributions from the
uncertainties due to chiral logarithms in the physical limit are
seen to be significantly less dominant. The mass difference
$\Delta M$ is $\sim 100$ MeV at the smaller quark masses, and
weakly dependent on $m_{\pi}^{2}$. The signature of repulsion at
quark masses near the physical regime would imply no evidence of
the resonance in the $J=2$ channel. If this mass difference from
two-$\phi$ threshold can be explained by the two-$\phi$
interaction, then the $s^{2}{\bar s}^{2}$ state can be regarded as
a two-$\phi$ scattering state.

\begin{table}[!h]
\caption{ \label{tab2} Hadron mass ratios at various pion masses
at $a_{s}=0.473$ fm. }
\begin{ruledtabular}
\begin{tabular}{ccccccccc}
$M_{\pi}(GeV)$ & $\frac{M_{4q}-2M_{\phi}}{M_{K}}$ &
$\big(\frac{M_{4q}-2M_{\phi}}{M_{K}}\big)^{*}$ & $\frac{M_{K}}{M_{\phi}}$\\
\hline
0.8249 & 0.223(3) &  0.296(5)  & 0.563(7) \\
0.6672 & 0.214(6) &  0.267(8)  & 0.523(8)\\
0.5015 & 0.209(8) &  0.252(10)  & 0.513(12)\\
0.4224 & 0.207(12) &  0.239(13)  & 0.512(14) \\
0.3617 & 0.193(15) &  0.247(18)  & 0.508(17)\\
0.2818 & 0.202(22) &  0.234(23)  & 0.510(22)\\
0.2218 & 0.193(26) &  0.232(27)  & 0.504(25)\\
\end{tabular}
\end{ruledtabular}
\end{table}

\begin{figure}[!h]
\scalebox{0.45}{\includegraphics{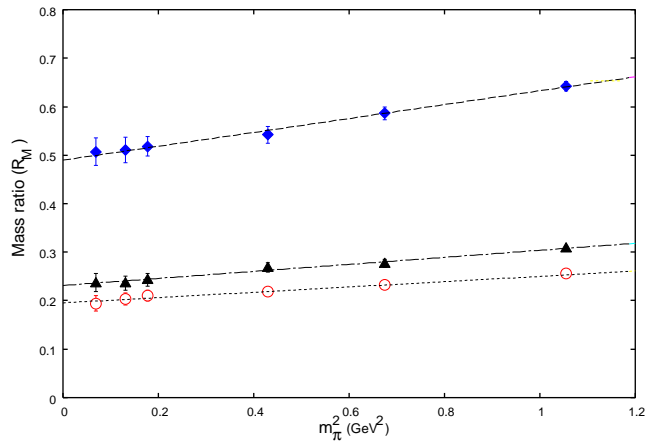}}
\caption{\label{fig5} As in Fig. \ref{fig4} but for larger quark
mass.}
\end{figure}

To verify whether analysis at relatively large quark masses would
affect the manifestation of the $J=2$ state and aid to confirm the
indication of a resonance, we  allow the quark mass to be
$m_{q}>2m_{s}$ so that the threshold for the decay $q^{2}{\bar
q}^{2}\rightarrow (q{\bar q})(q{\bar q})$ is elevated. The heavy
quark mass suppresses relativistic effects, which complicates the
interpretation of light-quark states. The resulting extracted mass
ratios are shown in Fig. \ref{fig5} and tabulated in Table
\ref{tab3}.

The behaviour observed for the mass differences between the $J=2$
and the two-particle states, at large quark masses, implies that
at larger quark masses, the data appear above the two-$\phi$
threshold by $\sim 95$ MeV and remains constant in magnitude as
the physical regime is approached. This trend continues in the
physical limit where the masses exhibit the opposite behaviour to
that which would be expected in the presence of binding. Again,
the positive mass difference could be a signature of repulsion in
this channel. This suggests that instead of a bound state, we
appear to be seeing a scattering state in $J=2$ channel. Since the
mass difference between the reported experimental $s^{2}{\bar
s}^{2}$ mass and the physical $2m_{\phi}$ continuum is $\sim 20$
MeV, the  observed signal is too heavy to be identified with the
empirical $f_{2}(1950)$ or $f_{2}(2010)$.
\begin{table}[!h]
\caption{ \label{tab3} Hadron mass ratios at larger quark
masses.}
\begin{ruledtabular}
\begin{tabular}{ccccccccc}
$M_{\pi}(GeV)$ & $\frac{M_{4q}-2M_{\phi}}{M_{K}}$ &
$\big(\frac{M_{4q}-2M_{\phi}}{M_{K}}\big)^{*}$ & $\frac{M_{K}}{M_{\phi}}$\\
\hline
1.2549 & 0.273(2) &  0.346(3)  & 0.764(9) \\
1.0272 & 0.256(4) &  0.308(5)  & 0.641(9)\\
0.8215 & 0.233(6) &  0.277(7)  & 0.586(11)\\
0.6552 & 0.219(7) &  0.269(9)  & 0.532(17) \\
0.4217 & 0.210(9) &  0.243(12)  & 0.5182(20)\\
0.3625 & 0.203(11) &  0.236(14)  & 0.5108(24)\\
0.2418 & 0.194(16) &  0.237(18)  & 0.5072(27)\\
\end{tabular}
\end{ruledtabular}
\end{table}

Using the physical kaon mass, $M_{K}=503(5)$ MeV, we obtain a mass
estimates of $2123(33)(58)$ MeV and $2137(39)(64)$ for the
$s^{2}{\bar s}^{2}$ tetraquark ground state and the second
ground-state, respectively. In each case, the first error is
statistical, and second one is our estimate of combined systematic
uncertainty including those coming from chiral extrapolation and
quenching effects. Note that  we cannot estimate the
discretization error since we have only one lattice spacing to
work with. Given the fact that the ratio does not show any scaling
violations, we could also quote the value of this quantity on our
finest lattice, which has the smallest error. Nevertheless, order
$2\%$ errors on the finally quoted values are mostly due to the
chiral extrapolations. The quenching errors might be the largest
source of uncertainty. Note however, that in the case of mass
ratios of stable hadrons, this is not expected to be very
important. It has been shown \cite{Gattringer03} that with an
appropriate definition of scale, the mass ratios of stable hadrons
are described correctly by the quenched approximation on the $
1-2\%$ level. To this end we also calculated the pseudoscalar to
vector meson ratio $R_{SP}$ and pseudoscalar to nucleon mass ratio
$R_{S N}$ and found that in the physical limit these ratios differ
about $1\%$ from their corresponding experimental values. So we
quote our quenching errors to be less than two percent.

\section{Summary and conclusion}

We presented the results of our investigation on the tetraquark
systems in improved anisotropic lattice QCD in the quenched
approximation. The mass of $J=2$ state was computed using field
operators, which are motivated by the non-$\pi\pi$ and diquark
structure. In the quenched approximation, our results suggest that
our interpolators have sufficient overlap with $f_{2}(ss{\bar
s}{\bar s})$ to allow a successful correlation matrix analysis and
produced the evidence that the mass of the lowest-lying state only
agrees marginally with the mass of $f(2010)$. In the region of
pion mass which we are able to access, we saw no evidence of
attraction that could be associated with the existence of a
resonance in $J=2$ channel.  Since our estimated value for the
mass of $f_{2}(s^{2}{\bar s}^{2})$ is marginally close to its
experimental value, we suspect that might be the $f(2010)$
resonance captured by our optimized correlator. However, on the
other hand, our spectral weight ratio for two different lattice
volumes deviates from one (the essential criterion for resonance)
with large errors for small quark masses, observed state exhibits
the expected volume dependence in the spectral weight for two
particles in a box.  The ground-state is found to be consistent
with scattering state. Our estimated values serve as predictions
of lattice QCD in quenched approximation. Indeed, our simulation
does not include dynamical quarks, the final conclusions will have
to wait till both disconnected correlators and annihilation
contributions are incorporated.

\begin{acknowledgments}
We are grateful for the access to the computing facility at the
Shenzhen University on 128 nodes of Deepsuper-21C. ML was
supported in part by the Guangdong Provincial Ministry of
Education.
\end{acknowledgments}

\end{document}